# Real-time control and monitoring system for LIPI's Public Cluster

## I. Firmansyah, B. Hermanto, Hadiyanto and L.T. Handoko

Group for Theoretical and Computational Physics, Research Center for Physics, Indonesian Institute of Sciences,
Kompleks Puspiptek Serpong, Tangerang 15310, Indonesia

**Abstract- We have developed a monitoring and control system for LIPI's Public Cluster. The system consists of microcontrollers and full web-based user interfaces for daily operation. It is argued that, due to its special natures, the cluster requires fully dedicated and self developed control and monitoring system. We discuss the implementation of using parallel port and dedicated micro-controller for this purpose. We also show that integrating such systems enables an autonomous control system based on the real time monitoring, for instance an autonomous power supply control based on the actual temperature, etc.**

## I. INTRODUCTION

Cluster computer is an infrastructure consisting of multiple nodes based on PC level computers working together to do either single or multiple numerical tasks simultaneously [1]. Beside the regular nodes, very large scale cluster has usually a control and monitoring system to maintain and monitor any system failures in the hardwares to prevent potential damages. On the other hand, the issue is not crucial in the small scale, namely less than 50 nodes, cluster facilities.

LIPI Public Cluster (LPC) is also a kind of such facilities, but it is open for public access for fully free [2,3]. In contrast with many regular cluster existing around the world which all users are well-known and provided with full access to the whole system as a block cluster to run their own jobs in particular period, in LPC the users could be anonymous and there might be multi users running varying jobs in different blocks of cluster under multi-block approach without any interferences among them [4]. This leads to a need for full online control and monitoring system over the internet although at time being it is still a small scale cluster. Therefore all different groups of users should have particular accesses to monitor and then control their own blocks of cluster.

The control system involves wide range of hardware's conditions, from turning each node on or off till automated control regarding the real-time conditions as (room, casing, processor, etc) temperatures and humidities. Then a real-time monitoring system for the whole hardware and also the detail of each of them is required. In order to realize this task properly, we have developed an integrated interface consisting micro-controllers and several censors.

We argue that this handy solution is flexible and simple, and finally the whole hardware is runned by web-based interface [5] to enable open and public access to realize the objectives of LPC [6].

In this paper we focus on the aspects of hardware based control and monitoring system in LPC. In the subsequent sections we first discuss the control and monitoring systems separately, followed with its integration and some discussions to summarize.

## II. CONNECTION (I/O) PORT

Before discussing the control and monitoring system, we should consider the way to connect the system with the targets, i.e. the I/O port for data acquisition and control system. Usually, most people tend to use a GPIB or data acquisition card due to its reliabilities and modularities, especially for those users who design a large scale and complex system. This option provides great benefit for users in simplifying the whole hardwares and in saving the design time as well, since there is no need to design everything from the scratch. Moreover one can further programme and interface it with the bundled softwares and its libraries for some popular low level languages as LabView or Visual Basic. All of these would enable creating attractive interfaces on the screen and store the acquired data with much less effort.

On the other hand, this approach is lack of flexibility and freedom, especially for those who need to realize special tasks that cannot be accomodated by the standard libraries. Of course ones could develop their own libraries to solve this problem. However this requires more efforts that might be comparable to developing the hardwares from the scratch. Also, there is a risk of high dependency on certain products which could bind the users forever.

Therefore, in our case we decide to utilize alternative connection port and develop the hardwares from the scratch to meet specific needs of LPC and also to reduce the overall cost as well. We have choosen the parallel port or LPT as the cheapest and easiest solution. Using the LPT, there is no need to design nor purchase expensive I/O port. We just need to plug the device we have developed into the port and programme it to be anything matches our purposes. Importantly, this can be used not only for control, but also for data acquisition depending on the programme.

Parallel port allows inputs of up to 9 bits or outputs of 12 bits at single time. The port consists of 4 control lines, 5 status lines and 8 data lines, and can be seen for instance in a PC as D-Type 25-pin female connector.

| Pin No (D-Type 25) | Pin No | SPP Signal | Direction In/out | Register |
|---|---|---|---|---|
| 1 | 1 | NStrobe | In/Out | Control |
| 2 | 2 | Data 0 | Out | Data |
| 3 | 3 | Data 1 | Out | Data |
| 4 | 4 | Data 2 | Out | Data |
| 5 | 5 | Data 3 | Out | Data |
| 6 | 6 | Data 4 | Out | Data |
| 7 | 7 | Data 5 | Out | Data |
| 8 | 8 | Data 6 | Out | Data |
| 9 | 9 | Data 7 | Out | Data |
| 10 | 10 | NAck | In | Status |
| 11 | 11 | Busy | In | Status |
| 12 | 12 | Paper Out | In | Status |
| 13 | 13 | Select | In | Status |
| 14 | 14 | nAuto-Linefeed | In/Out | Control |
| 15 | 32 | nError / nFault | In | Status |
| 16 | 31 | nInitialize | In/Out | Control |
| 17 | 36 | nSelect-Printer / nSelect-In | In/Out | Control |
| 18 - 25 | 19 - 30 | Ground | | Gnd |

Tab. 1 shows the assignment of the "Pin Outs" of D-Type 25-pin connector and Centronics 34-pin connector. Just to mention, there are many ways on how to assign the I/O ports in many programming languages. For example, in C language we can use I/O address 378h as follows :

*outp(0x378,N);   or  outportb(0x378,N);*

to send the data *N* out [7].

Thus this enables, for instance controlling only up to 8 devices simultaneously. Since in LPC we have, in principle unlimited number of nodes as independent devices to control and monitor simultaneously, it is clear that we should develop an additional microcontroller based device between the I/O port and the nodes [8]. The microcontroller then acts as if a multiplexer or demultiplexer as will be explained soon in the subsequent sections.

Now we are ready to discuss how we control and monitor the nodes in LPC through parallel port and by utilizing the microcontrollers.

## III. CONTROL SYSTEM

The main purpose of control system in LPC is to turn on or to shut down the main power supply of each node independently, either manually or automatically according to certain conditions. As mentioned above, the microcontroller acts as if a demultiplexer. In the sense of control system, it works by getting certain command from the parallel port and then decides current condition of each node to execute certain task on it.

We realize this purpose by assembling a relay for each node. However since the current source from parallel port is too low to drive the relay properly, we need to amplify the current out of parallel port. This can be done by bringing transistor as the relay driver by activating it in saturation condition. In our case we have deployed the easier way using current amplifier IC such as ULN2803. We could generally benefit from using this IC, in particular when more outputs are needed. In fact it will also simplify the circuits as shown in Fig. 1.

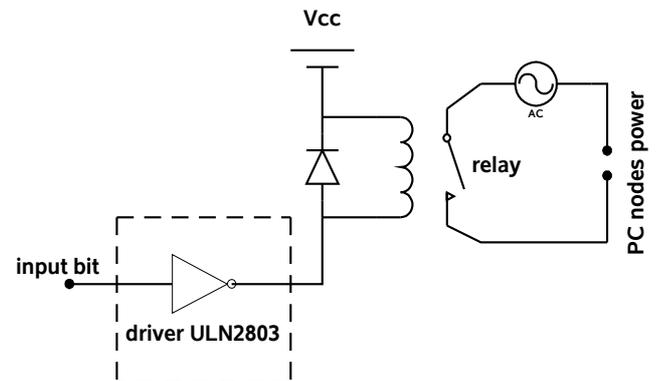

Fig. 1. Relay driver using ULN2803 in the control system.

Further we have developed an algorithm for microcontroller to send the command to control the relays through parallel port. As reference, below is the flow of such algorithms : [9,10]

```
#include <xxx.h>          //depending on microcontroller used
void main()
{
//input port for capturing the command sent by parallel port
//can be set depending on capabilities of microcontroller
    while(1)          //loop forever
    {
    data=input();     //read command into variable data
    switch(data)
        {
        case '1' : turn on node1;
        case '2' : turn on node2;
        …………
        case '129' : turn off node1;
        case '130' : turn off node2;
        ………….
        }}
}
```

## IV. MONITORING SYSTEM

Similar prescription is valid for the monitoring system to acquire data from certain censors through the microcontroller and parallel port. In LPC, we have embedded the temperature and humidity censors in each row (that is a single huge casing) of clusters to capture the actual conditions inside the casing, beside the internal temperature of each processor which can be read easily through the BIOS.

In contrast with the control system, in this case the microcontroller acts as a multiplexer. It selects the inputs from multiple censors. In order to simplify the circuits we have implemented a microcontroller with built-in ADC. That means we do need to use an additional external ADC anymore, actually we use Atmega8535L from Atmel [11].

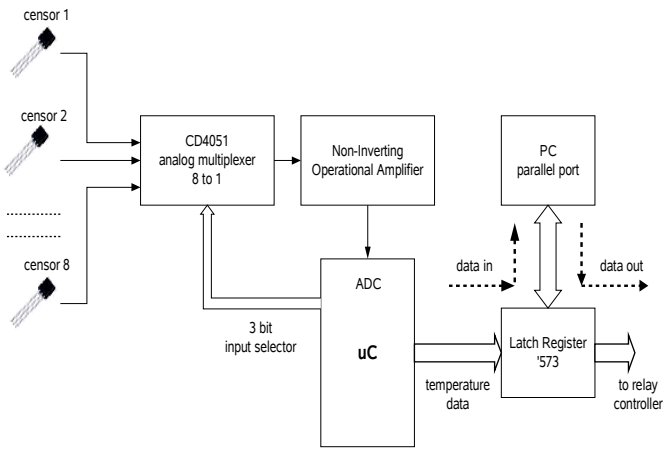

Fig. 2. Block diagram of the monitoring system.

As the data acquisition is over parallel port which is used for controlling, we have to make sure that the port could be used in two directions simultaneously. This can be done by implementing two latch registered IC such as 74LS573 to make the parallel port accessible from two directions [11]. This is described in Fig. 2.

To read some inputs from multiple censors, first the input selector or multiplexer is needed to select the data source. We use LM35 which could produce output voltage 10mV/$^o$ C, in which the output voltage is 0 V at 0 $^o$ C and 1 V at 100 $^o$ C. The analog multiplexer CD4051 is used here to select which LM35 will be processed where the input address in this IC is controlled by microcontroller. CD4051 has eight signals input, one is output signal and gives possibility to choose which input is connected to the output. The non-inverting amplifier then is used to amplify the signal before being converted into digital data by ADC0804 [12]. The last section is data processing by microcontroller itself so that the final result could be sent to PC parallel port as a data of particular node's temperature.

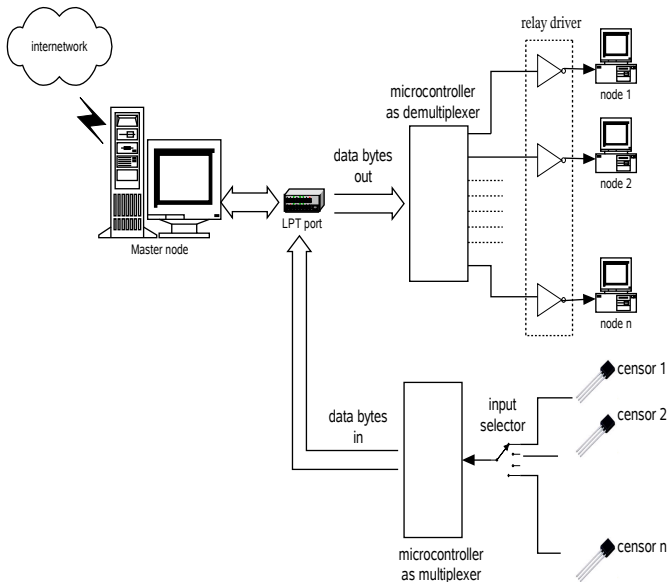

Fig. 3. Block diagram of the integrated control and monitoring system.

As done in the control system, we have deployed a code for microcontroller to select one from multiple censors as below : [13]

```
// code sniffed to select input censor
#include <xxx.h>          // depend on microcontroller used
void main()
{
while(1)          //loop forever
        {
        data=read_command();     //read command
        switch(data)
        {
         case '1':
                select censor1 ;     // setting input address
                process data;        // activating ADC
                send data;           //send data to PC
         case '2':
                select censor2 ;     // setting input address
                process data;        // activating ADC
                send data;           //send data to PC
                .............
                .............
        }}
}
```

## V. SUMMARY

Finally, we have developed a control and monitoring system for LPC. The number of node that can be controlled is presently 48 following the initial design of LPC. However the extension is quite trivial and can be done soon according to the expansion planning in the future.

The integration of both hardware systems has been done in a compact dedicated casing with centralized power-supply and connection over a single parallel port. The whole system is described in Fig. 3. In this design we use two microcontrollers mcs51 family from Atmel, in which the input data from parallel port is connected to port1(P1) of each microcontroller, and the rest port (P0, P2, P3) is used as output data to drive a relay. Beside that, the system also measures the casing's temperature using LM35 censors.

The web interface is used to easily monitor and control the whole system over web. We should emphasize that though, for instance the casings' temperatures are acquired once per ten minutes, this delay time is small enough in the sense of real needs in monitoring LPC. So in that sense we can say that we have deployed a real-time monitoring system in LPC. The integration of both systems enables automatic control based on the real-time monitoring results, that is very crucial in a full-time running cluster to prevent further damages due to system failures. The current system also enables automatic partial activation / deactivation of each node depending on its usages wchich can reduce the overall running cost.

We should also note that the current integrated web-based interface involves automatic decision making based on the Extended Genetic Algorithm to obtain the most optimum set of nodes for each incoming request [14]. Combining this algorithm with the current control system and web-based interface, we are going to activate full automatic allocation resources to enable an instant approval and allocation once a new request has been submitted.

## ACKNOWLEDGMENT

This work is financially supported by the Riset Kompetitif LIPI in fiscal year 2007 under Contract no. 11.04/SK/KPPI/II/2007 and the Indonesia Toray Science Foundation Research Grant 2007.